\documentstyle[seceq,supplement,epsf]{ptptex}

\def\phi{\varphi}
\def\half{\mbox{\small $\frac{1}{2}$}}
\def\ltap{\raisebox{-.4ex}{\rlap{$\sim$}} \raisebox{.4ex}{$<$}}

\preprintnumber[3cm]{SHEP 98-??\\ hep-th/9801???}

\title{
Elements of the Continuous Renormalization Group
}

\author{ Tim R. {\sc Morris}
}

\inst{
Physics Department,\\ University of Southampton,
\\ Highfield \\ Southampton, SO17 1BJ\\
U.K.
}

\recdate{
\today
}

\abst{%
These two lectures cover some of the advances that underpin
recent progress in deriving continuum solutions from the
 exact renormalization group. We concentrate on concepts
and on exact non-perturbative statements, but in
the process will describe how real non-perturbative calculations can
be done, particularly within derivative expansion approximations.
An effort has been made to keep the lectures pedagogical and self-contained.
Topics covered are the derivation of the flow equations, their equivalence,
continuum limits, perturbation theory, truncations, derivative expansions,
identification of fixed points and eigenoperators, and 
the r\^ole of reparametrization invariance.
Some new material is included, in particular a demonstration of
non-perturbative renormalisability, and a discussion
of ultraviolet renormalons. 
}

\begin{document}

\maketitle

\section{Introduction}
As stated above, these lectures will
 concentrate on exact statements, the conceptual advances,
in the  exact renormalization group, a.k.a.
Wilson's continuous renormalization group:\cite{kogwil}
This is motivated by the  belief
that these are ultimately the  most important 
aspects of the recent progress,  but 
at the same time this viewpoint lends itself to a (hopefully)
elegant and pedagogical introduction to this area.
This means  however 
that  applications will not be reviewed, or practical matters such as
the accuracy of 
approximations discussed {\it per se}.
 Individuals interested to learn more about these issues, 
are encouraged to consult our reviews\cite{zak,rg96} and the lectures
by Aoki and Wetterich in this volume. 
Suffice to say here that there {\sl are}
approximations, in particular the derivative expansion,
 which give fair to accurate numerical results in practice. 
The motivation fueling the recent progress is the need
to derive better analytic
approximation methods for truly non-perturbative quantum field theory,
i.e. where there are no small parameters\footnote{Typical small parameters
that are sometimes useful are small coupling, i.e. perturbation theory, or
$1/N$ where $N$ is the number of components of a field, or $\epsilon=4-D$
where $D$ is the space-time dimension.}
one can fruitfully expand in. There is a clear need for such approaches,
of course within the archetypical example -- low energy QCD, 
but perhaps more importantly 
in the need to better understand (even qualitatively)
the possibilities offered by the full parameter space 
of non-perturbative quantum field theories, such as
 may explain some of the mysteries of the symmetry breaking sector of
the Standard Model (for example), and/or Planck scale physics. 
On the other hand, the issue of renormalisability, which in many approaches
appears as a subtle problem -- particularly so for approximations that
do not rely on expansion in some small parameter\cite{erg}, is
essentially trivial within the exact Renormalization Group (RG), 
as we will see. This means that within the framework of the continuous
RG, almost {\sl all} approximations
preserve a crucial property of quantum field theory, namely the existence
of a continuum limit. Moreover, as we will see,  this
framework allows us to find systematically\footnote{in some
approximation scheme, e.g. derivative expansion} 
{\sl all possible} non-perturbative continuum limits
 within the {\sl infinite dimensional} 
parameter space of all possible quantum field theories
with a given field content and symmetries.

As mentioned briefly in the abstract, the topics covered are as follows.
In sect. 2, we cover the derivation of the exact RG flow equations,
Polchinski's version and the Legendre flow equation, in such a way
that it is clear that they correspond to integrating out modes and
that they are equivalent to each other. In particular we emphasize the
important consequence that the Green functions of the theory may be
extracted directly from the Wilsonian effective action. 
In sect.3, we
show how solutions for the effective action corresponding to continuum
limits may be accessed directly in renormalised terms, and sketch a
{non-perturbative} proof of self-similarity (renormalisability)
of continuum solutions.  We delineate the r\^ole of fixed points,
eigenperturbations and renormalised trajectories.  We show that the
direct solution in renormalised variables follows particularly simply in
perturbation theory via a process of iteration.  We use this to discuss
the (non)existence of the continuum limit of four dimensional scalar
field theory and show how this is related to the appearance of ultraviolet
renormalons, whose existence follows very naturally in this formalism.
In sect.4, we briefly cover non-perturbative approximations, noting
that these preserve renormalisability. In sects. 5 and 6, we use the
local potential approximation  to show
how fixed points are determined by the requirement that they are non-singular,
and eigenperturbations are determined through the requirement of 
self-similarity.
Finally, in sect.7, we show how the fields anomalous dimension is determined
through the property of reparametrization invariance. We demonstrate the
existence of this symmetry in a simple example (the Gaussian fixed point),
and explain how it may be
maintained in derivative expansions, and
the problem that arises if it is broken. 

\section{The RG flow equations}
The basic idea behind the (continuous) RG is illustrated in fig.1. 
\begin{figure}[ht]  
\epsfxsize=0.6\hsize
\hfil\epsfbox{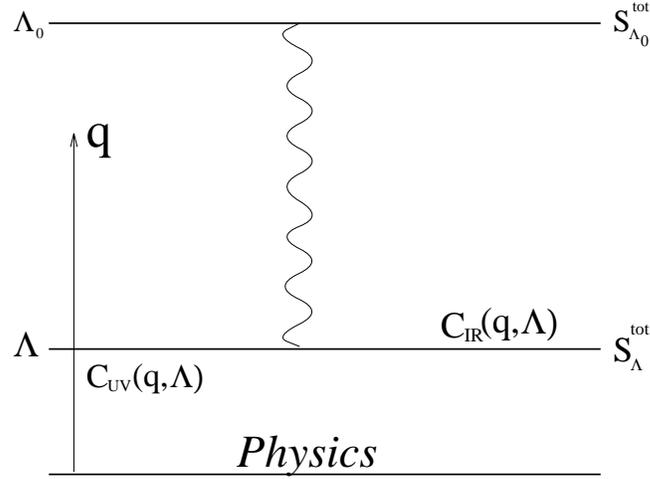}\hfill
\caption{Integrating out with the continuous RG.}
\end{figure}
Rather than integrate over all momentum
modes $q$ in one go, one first 
integrates out 
modes  between a cutoff scale $\Lambda_0$ and a very much
lower energy scale $\Lambda$.
Both of these scales are introduced by hand. 
The remaining integral from $\Lambda$ to zero may again be expressed as
a partition function, but the bare action $S^{tot}_{\Lambda_0}$ (which is
typically chosen to be as simple a functional as possible) is replaced
by a complicated effective action $S^{tot}_{\Lambda}$  and the
overall cutoff $\Lambda_0$ by the effective cutoff $\Lambda$, in such a
way that all physics i.e. all  Green functions, are left invariant.
It may seem at first sight that such a partial integration step merely
complicates the issue. For example, we have had to replace the (generally)
simple $S^{tot}_{\Lambda_0}$ by a complicated $S^{tot}_{\Lambda}$. However,
for the most part the complicated nature of $S^{tot}_{\Lambda}$ 
merely expresses
the fact that quantum field theory itself
is complicated. Indeed from fig.1, we see that we can regard 
the cutoff $\Lambda$  as an infrared cutoff for the
 modes $q$ that have already been integrated out. Thus we should expect that
$\exp-S^{tot}_{\Lambda}$ is (more or less)
the original partition function for the quantum field theory, but
modified by an infrared cutoff $\Lambda$.
This means in particular,
 that we can recover all Green functions of the theory
from  $\lim_{\Lambda\to0}S^{tot}_{\Lambda}$. This statement is 
surprising if one adheres to the view
that Wilsonian RG steps involve a loss of information (and thus 
a complete loss of information when $\Lambda\to0$),
but it lies at the heart of why the present techniques allow valuable
approximation methods for quantum field theory. Therefore
let us sketch 
a proof\cite{erg}. (Somewhat similar statements 
have appeared elsewhere\cite{oerg}\cite{wet}.)

We will introduce the effective ultraviolet cutoff by modifying
propagators $ 1/q^2$ to $\Delta_{UV}=
C_{UV}(q,\Lambda)/q^2$, where $C_{UV}$
is a profile that acts as an ultra-violet cutoff\cite{pol},
i.e. $C_{UV}(0,\Lambda)=1$ and 
$C_{UV}\to0$ (sufficiently fast) as $q\to\infty$.
Similarly, we introduce an infrared cutoff by  modifying
propagators $ 1/q^2$ to $\Delta_{IR}=
C_{IR}(q,\Lambda)/q^2$, where $C_{IR}$ is a profile with the 
properties  $C_{IR}(0,\Lambda)=0$ and $C_{IR}\to1$ as $q\to\infty$.
We require that the two cutoffs are related as follows,
\begin{equation}
\label{sum}
C_{IR}(q,\Lambda)+C_{UV}(q,\Lambda)=1\quad.
\end{equation}
Then we have the identity (up to a constant of
proportionality\footnote{From now on we will drop  
these uninteresting constants of proportionality.}),\cite{erg}
\begin{eqnarray}
\label{zorig} Z[J]& &= \int\!{\cal D}\phi\
\exp\{-\half\phi\cdot q^2\cdot\phi-S_{\Lambda_0}[\phi]+J\cdot\phi\}\\
\label{grle} & &=\int{\cal D}\phi_{>}{\cal D}\phi_{<}\
 \exp\{-\half\phi_{>}\cdot\Delta_{IR}^{-1}\cdot\phi_{>}
 -\half\phi_{<}\cdot\Delta_{UV}^{-1}\cdot\phi_{<} \\
 &&\nonumber\phantom{=\int{\cal D}\phi_{>}{\cal D}\phi_{<}\ \exp\{ }
-S_{\Lambda_0}[\phi_{>}+\phi_{<}]+J.(\phi_{>}+\phi_{<})\}\quad,
\end{eqnarray}
where $S_{\Lambda_0}$ is the interaction part of the bare action.
In view of their respective propagators, we interpret the $\phi_{>}$ 
field as the momentum
modes higher than $\Lambda$, and the $\phi_{<}$ field
as the modes that are lower than $\Lambda$. [But please note that
the truth is fuzzier unless the cutoff is $C_{UV}=\theta(\Lambda-q)$, 
i.e. sharp.
When the cutoff is smooth, modes lower (higher)
than $\Lambda$ in $\phi_{>}$ ($\phi_{<}$)
are only damped.]

To see that the identity (\ref{grle}) is true perturbatively, note that
(as illustrated in fig.2) as a consequence of the sum form in the interactions,
every Feynman diagram constructed from (\ref{zorig}) now appears twice
for every internal propagator it contains: once with $1/q^2$ replaced by
$\Delta_{UV}$ and once with $1/q^2$ replaced by $\Delta_{IR}$. 
Thus for every such
propagator line, what actually counts is the sum, which is 
however just $1/q^2$ again, by (\ref{sum}).
The non-perturbative proof is almost as trivial: one simply makes some
shifts on the fields and performs a Gaussian integration.\cite{erg}
\begin{figure}[ht]  
\epsfxsize=0.7\hsize
\hfil\epsfbox{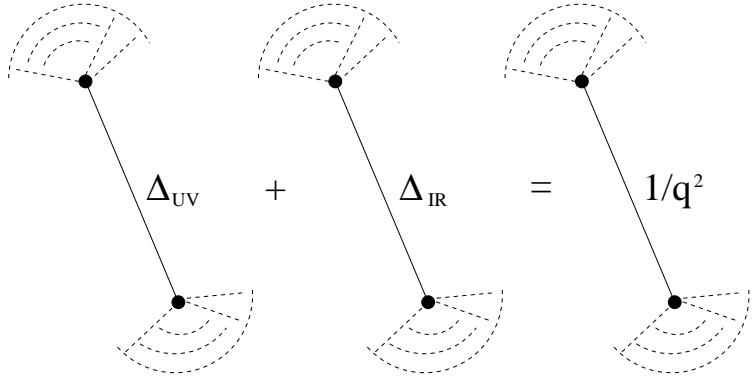}\hfill
\caption{ Feynman diagram representation of the identity (\ref{grle}).}
\end{figure}

Now consider only integrating over the higher modes:
\begin{equation}
\label{horig}
Z_\Lambda[J,\phi_{<}] =\int{\cal D}\phi_{>}\
 \exp\{-\half\phi_{>}\cdot\Delta_{IR}^{-1}\cdot\phi_{>}
 -S_{\Lambda_0}[\phi_{>}+\phi_{<}]+J.(\phi_{>}+\phi_{<})\}\quad.
\end{equation}
By a similar shift of variables, it is straightforward to show that 
$Z_\Lambda$ does not depend on both $J$ and $\phi_{<}$ independently, but
essentially only on the sum 
\begin{equation}\label{phi}
\phi=\Delta_{IR}\cdot J +\phi_{<}\quad.
\end{equation} 
The exact statement is\cite{erg}
\begin{equation}
\label{S}
Z_\Lambda[J,\phi_<]=
\exp\left\{ \half J\cdot\Delta_{IR}\cdot J +J\cdot\phi_<
-S_\Lambda[\Delta_{IR}\cdot J +\phi_{<}]\right\}\quad,
\end{equation}
for some functional $S_\Lambda$. What is the meaning of $S_\Lambda$?
In contrast to some works\cite{pol}
 we have not restricted the support of $J$ to low energy modes only.
Had we done so, we would have had $\Delta_{IR}\cdot J=0$. In this
case (\ref{S}) simplifies, but from (\ref{horig}) and (\ref{grle}),
\begin{equation}
Z[J]=\int\!\!{\cal D}\phi_{<}\ Z_\Lambda[J,\phi_<]\,
 \exp\{ -\half\phi_{<}\cdot\Delta_{UV}^{-1}\cdot\phi_{<} \}\quad.
\end{equation}
We see that $S_\Lambda$ is nothing but the interaction part of the
Wilsonian effective action $S_\Lambda^{tot}$. This is a nice result.
There is no price to pay for letting $J$ couple to all modes: although the
effective dependence on $J$ is now non-linear, its dependence is no worse
than that already contained in $S_\Lambda$.  As we see from (\ref{phi}),
the dependence on $J$, is essentially carried by the higher modes of
$\phi$.

The exact RG equations, or `flow equations', follow readily from
the fact that (\ref{horig}) depends on $\Lambda$ only through the
$\phi_{>}\cdot\Delta_{IR}^{-1}\cdot\phi_{>}$ term.\cite{erg}
 Thus, differentiating
$Z_\Lambda$ with respect to $\Lambda$ we  obtain
immediately the flow equation for $Z_\Lambda$:
\begin{equation}
\label{Zfl}
{\partial\over \partial\Lambda} Z_\Lambda[\phi_<,J] =-{1\over2}
\left({\delta\over\delta J}-\phi_<\right).\left({\partial\over
\partial\Lambda}\Delta_{IR}^{-1}\right).\left({\delta\over\delta J}-
\phi_<\right)\
Z_\Lambda\quad.
\end{equation}
And substituting (\ref{S}), yields
Polchinski's version\cite{pol} of Wilson's flow equation:\cite{kogwil}
\begin{equation}
\label{Sfl}
{\partial\over\partial\Lambda}  S_{\Lambda}[\phi]
={1\over2}{\delta S_{\Lambda}\over\delta\varphi}\cdot{\partial\Delta_{UV}\over
\partial\Lambda}\cdot{\delta S_{\Lambda}\over\delta\varphi}-{1\over2}{\rm tr}
{\partial\Delta_{UV}\over\partial\Lambda}\cdot{\delta^2 S_{\Lambda}\over
\delta\varphi\delta\varphi}\quad.
\label{Pfl}
\end{equation}
(Of course here $\partial/\partial\Lambda$ is to be taken at constant $\phi$.)
On the other hand, if we recognize (\ref{horig}) as a partition function
for an infrared cutoff theory (in an `external' field $\phi_<$), 
we can by the standard formulae, construct the Legendre effective action
$\Gamma^{tot}_\Lambda[\phi^c,\phi_<]$. Here $\phi^c$ is the classical
field (defined in the usual way by  $\phi^c=\delta \ln Z_\Lambda/\delta J$).
The fact that $Z_\Lambda$'s dependence is prescribed through (\ref{S}),
turns out to imply that $\Gamma^{tot}$'s dependence on $\phi_<$ is very simple:
\begin{equation}
\Gamma^{tot}_\Lambda[\phi^c,\phi_<]=\half(\phi^c-\phi_<)
\cdot\Delta_{IR}^{-1}
\cdot(\phi^c-\phi_<)+\Gamma_\Lambda[\phi^c]\quad.
\end{equation}
$\Gamma_\Lambda$ contains the effective interactions of the Legendre effective
action. Substituting the Legendre transform equations into (\ref{Zfl}), 
one readily obtains its flow equation\cite{erg,wet,legfl,deriv}
\begin{equation}
\label{Gfl}
{\partial\over\partial\Lambda}\Gamma_\Lambda[\phi^c]=
-{1\over2}{\rm tr}\left[{1\over\Delta_{IR}}{\partial\Delta_{IR}\over\partial
\Lambda}\cdot\left(1+\Delta_{IR}\cdot{\delta^2\Gamma_\Lambda\over
\delta\varphi^c\delta\varphi^c}\right)^{-1}\right]\quad.
\end{equation}
From (\ref{S}), it is clear that $S_\Lambda$ is (essentially) the
generator of connected Green functions in the infrared cutoff theory.
Therefore, there is a Legendre transform relation that maps between
the Wilsonian effective action and (infrared cutoff) Legendre effective 
action:\cite{erg}
\begin{equation}
\label{legtr}
S_\Lambda[\phi]=\Gamma_\Lambda[\phi^c]+{1\over2}(\phi^c-\phi)\cdot
\Delta^{-1}_{IR}\cdot(\phi^c-\phi)\quad.
\end{equation} 
We see that the $\Lambda\to0$
limit of the Wilsonian effective action can be related to the standard
Legendre effective action through $\Gamma[\phi^c]=\lim_{\Lambda\to0}
\Gamma_\Lambda[\phi^c]$, and hence to Green functions, S matrices,
classical effective potentials, and so forth. On the other hand,
since the infrared cutoff Legendre effective action is just a Legendre
transform of the
Wilsonian effective action, we can expect it also to have 
fixed point and self-similar RG behaviour. 

To see this RG behaviour however, it is necessary to add the other essential ingredient of an RG
`blocking' step:
scaling the cutoff  back to its original size. Simpler and equivalent, is
to ensure that all variables are `measured' in units of $\Lambda$, 
i.e. we change variables to
ones that are dimensionless, by
dividing by $\Lambda$ raised to the power
 of their scaling dimensions.\cite{rg96}
From now on, we will assume that this has been done.

\section{Continuum limits}
This RG method of describing quantum field theory becomes advantageous when
we consider the continuum limit.  
We will indicate how  one can solve the flow equations in this case,
directly in the continuum, dispensing with the standard, but for quantum 
field theory, actually artificial and extraneous, scaffolding of imposing
an overall cutoff $\Lambda_0$, finding a sufficiently general bare action
$S_{\Lambda_0}$, and then tuning to a continuum limit as $\Lambda_0\to\infty$.

The simplest case corresponds to a fixed point of the flow, $S_\Lambda=
S_*$ with
\begin{equation}
\label{fp}
\Lambda{\partial\over\partial\Lambda}S_*[\phi]=0\quad.
\end{equation}
(We will write $S_\Lambda$ here and later,
but the same comments apply for its Legendre
transform $\Gamma_\Lambda$.)
Because all dimensionful variables have been exchanged for dimensionless
ones using $\Lambda$, independence of $\Lambda$
 implies that $S_*$ depends
on no scale at all, i.e. corresponds to physics that is scale invariant,
and thus in particular describes a massless continuum limit. (You can
see that it must be massless, for otherwise the mass would set the scale.
On the other hand, it must be a continuum limit, corresponding to 
$\Lambda_0\to\infty$, for otherwise $\Lambda_0$ would set the scale.)
{\it Nota bene}, massless continuum limits thus follow directly from
fixed points (\ref{fp}); no tuning or bare actions required!

In the massive case, as a consequence of the fact that the flow equations
are sensitive only to momenta of order $\Lambda$, we can obtain 
the continuum solutions
{\sl directly} in terms of the renormalized  variables, viz. the
renormalised field $\phi$,\cite{eqs}\footnote{i.e. scaled to give a normalised
kinetic term} the relevant\footnote{masses are included as
one of the relevant couplings}
and marginally-relevant couplings, say $g^1$ to $g^n$, and the
anomalous dimension  (wavefunction renormalization) $\gamma$:
\begin{equation}
\label{ssfl}
S_\Lambda[\phi]=S[\phi]\left(g^1(\Lambda),\cdots,g^n(\Lambda),
\gamma(\Lambda)\right)\quad.
\end{equation}
Note that all the scale dependence appears only in the couplings and $\gamma$.
This {\sl self-similar evolution} is equivalent to the statement of
renormalisability (because it shows that $S_\Lambda$ does not explicitly 
depend on $\Lambda_0/\Lambda$). Let us show, as advertised, that this 
property, the subject of long and subtle perturbative proofs, follows
essentially trivially in this framework, even non-perturbatively. 

\subsection{Renormalisability}
 Before doing this
however, we need to recall the standard lore\cite{kogwil} on how a 
non-perturbative  
massive continuum limit is obtained in this framework. 
This is illustrated in fig.3. 
\begin{figure}[ht]  
\epsfxsize=0.6\hsize
\hfil\epsfbox{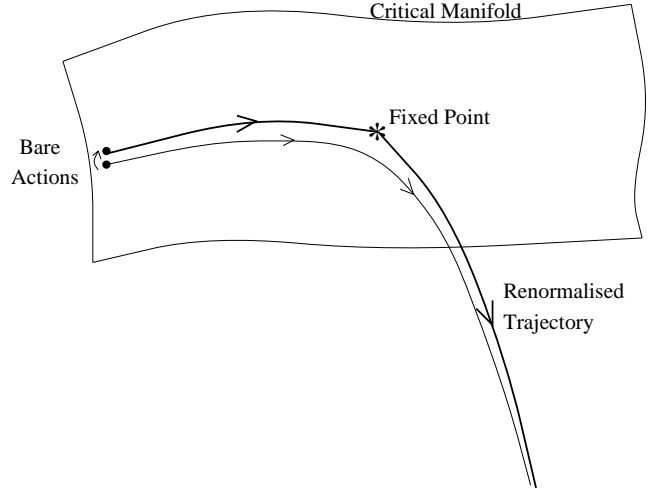}\hfill
\caption{Tuning to a massive continuum limit.}
\end{figure}
In the infinite dimensional space of bare actions, there is the
so-called 
critical manifold, which consists of all bare actions yielding a given
massless continuum limit. Any point on this manifold -- i.e. any such
bare action -- flows under a given RG towards its fixed point; local
to the fixed point, the critical manifold is spanned by the infinite set
of irrelevant operators. The other directions
emanating out of the critical manifold at the fixed point,
are spanned by relevant 
and marginally relevant perturbations
(with RG eigenvalues $\lambda_i>0$ and $\lambda_i=0$,
respectively).
Choosing an appropriate parametrization of the bare action, we move a little
bit away from the critical manifold. The trajectory of the RG will 
to begin with, move towards the fixed point, but then shoot away along one of
the relevant directions towards the so-called
high temperature fixed point which represents an infinitely massive 
quantum field theory. 

To obtain the continuum limit, and thus finite
masses, one must now tune the bare action back towards the critical
manifold and at the same time, reexpress
 physical quantities  in renormalised terms
appropriate for the diverging correlation length. In the limit that the
bare action touches the critical manifold, the RG trajectory splits into
two: a part that goes right into
the fixed point, and a second part that emanates out from the fixed point
 along the relevant directions. This path
is known as a Renormalised Trajectory\cite{kogwil} (RT). The effective
actions on this path are `perfect actions'.\cite{H} In terms of renormalised
quantities, the far end
of this path obtains a finite limit,
namely the effective action of the continuum quantum field theory.

Therefore, to obtain the massive continuum limit directly we must first
describe the RT. 
Clearly from the above discussion, the RT is fixed by
specifying that it emanates from the fixed point and giving the `rates'
in the relevant and marginally relevant directions:
\begin{equation}
\label{RTbc}
S_\Lambda=S_*[\phi]+\sum_{i=1}^n \alpha^i (\mu/\Lambda)^{\lambda_i}
{\cal O}_i[\phi]\quad\quad\quad
{\rm as} \quad \Lambda\to\infty\quad.
\end{equation}
Here the ${\cal O}_i[\phi]$ are the eigenperturbations 
conjugate to the couplings $g^i$, the $\alpha^i$ are integration constants
-- the `rates',
(which should be taken finite\cite{eqs}) and  
$\mu$ is as usual, an arbitrary finite mass scale.
Since the RG equations are first-order in 
$\Lambda$, it is sufficient for a trajectory
to give a  
boundary value for $S_\Lambda$ (at some point $\Lambda$) except
 at a  fixed point, which is a so-called singular point for the differential
equation. Here  this slightly more 
subtle boundary condition is required.
 The power law behaviour in $\Lambda$ follows
simply from expansion of the flow equations (\ref{Sfl},\ref{Gfl})
to first order in the perturbation, and by separation of variables

(Actually, for marginal perturbations, $\lambda=0$, it is necessary
to follow the evolution to second order and the power law behaviour
in (\ref{RTbc}) is replaced by logarithmic evolution. For marginally
relevant perturbations the multiplying factor still decays back
into the fixed point as $\Lambda\to\infty$. We will ignore these minor
complications in this subsection and persist
with  formulae appropriate for the strictly
relevant cases $\lambda>0$. There is however a more subtle issue
buried in our assumption that the perturbations can be treated to
first order.
Since (\ref{RTbc}) incorporates a limit as $\Lambda\to\infty$,
 this
assumption looks innocent enough for $\lambda>0$, but we will see later
that it is true only
for certain quantized perturbations.)

The boundary condition (\ref{RTbc}) and the RG flow equation, completely
specify the massive continuum limit, i.e. the 
continuum limit is fully specified as
\begin{equation}
\label{Sa}
S_\Lambda[\phi]\equiv S_\Lambda[\phi]( \alpha^1,\cdots,\alpha^n)\quad.
\end{equation}
Once again, this is achieved 
directly -- without $\Lambda_0$, bare actions or tuning. 
Let us define the renormalised couplings $g^i(\Lambda)$ such that 
\begin{equation}
\label{gbc}
 g^i\sim\alpha^i(\mu/\Lambda)^{\lambda_i}\quad\quad\quad
{\rm as} \quad \Lambda\to\infty\quad,
\end{equation}
i.e. define renormalization conditions consistent with the form of the
${\cal O}_i[\phi]$. 
Evidently, by applying the renormalization conditions directly to (\ref{Sa})
we may explicitly
read off the renormalized couplings:  
$g^i\equiv g^i({\underline\alpha},\Lambda)$.
[For example in four dimensional
$\lambda\phi^4$ theory, the renormalization condition might
be to define $\lambda(\Lambda)$ as the value of the four point function
at zero momentum. Reading this value off from (\ref{Sa}) then
directly defines $\lambda({\underline\alpha},\Lambda)$.] 
But given the functions $g^i(\Lambda)$ and the values of the couplings,
we can invert to find $\Lambda$, and the $\alpha^i=\lim_{\Lambda\to\infty}
(\Lambda/\mu)^{\lambda_i}g^i(\Lambda)$.
 Therefore  the couplings ${\underline g}(\Lambda)$ [and
$\gamma(\Lambda)$] provide entirely equivalent 
information for specifying the solution as the ${\underline\alpha}$
and $\Lambda$. Exchanging the latter for the former in (\ref{Sa}), gives
(\ref{ssfl}), and renormalisability is thus shown, as required.

Using the renormalization conditions on
the flow equation (\ref{Sfl}) (which scaled, does not depend explicitly on $\Lambda$),
we read off from the left hand side the functions $\beta^i=\Lambda
\partial g^i/\partial\Lambda$, and from the right hand sides 
explicit non-perturbative
expressions for these $\beta^i\equiv\beta^i({\underline g},\gamma)$.
This self-similar form follows directly from (\ref{ssfl}).

\subsection{Perturbation Theory}
In perturbation theory, 
this direct continuum solution of the flow equations,
 follows particularly simply, by iteration. This was demonstrated
in a model approximation of four dimensional $\lambda \phi^4$ 
theory\cite{erg} and it will serve here as a simple illustration.
To construct the model we expanded both sides of the flow equation in
 $\phi$ and momenta, and kept only the coefficients in
front of $\phi^4$ and a certain irrelevant operator:
\begin{eqnarray}
\label{lams}
\beta=\Lambda{\partial\lambda\over\partial\Lambda} &&= 
{3\over(4\pi)^2}(\lambda+2\gamma_1)^2 \\
\label{gams}
\Lambda{\partial\gamma_1\over\partial\Lambda} &&=
\gamma_1-{1\over24\pi^3}(\lambda+2\gamma_1)(\lambda-[3\pi+2]\gamma_1)\quad.
\end{eqnarray}
Here $\lambda(\Lambda)$ is the four point coupling, and $\gamma_1(\Lambda)$
the irrelevant coupling. (The irrelevant operator's precise form is
immaterial for the present purposes. It arises in the momentum expansion
of the sharp cutoff flow equations and is responsible for 99\% of the
two-loop $\beta$ function coefficient.\cite{erg,truncm})

Solving these equations numerically, starting with 
$\lambda(\Lambda_0)=\lambda_0$ and $\gamma_1(\Lambda_0)=0$,
 results in the curves shown in fig.4.\cite{erg}
\begin{figure}[ht] 
\vskip -.5cm 
\epsfxsize=0.8\hsize
\hfil\epsfbox{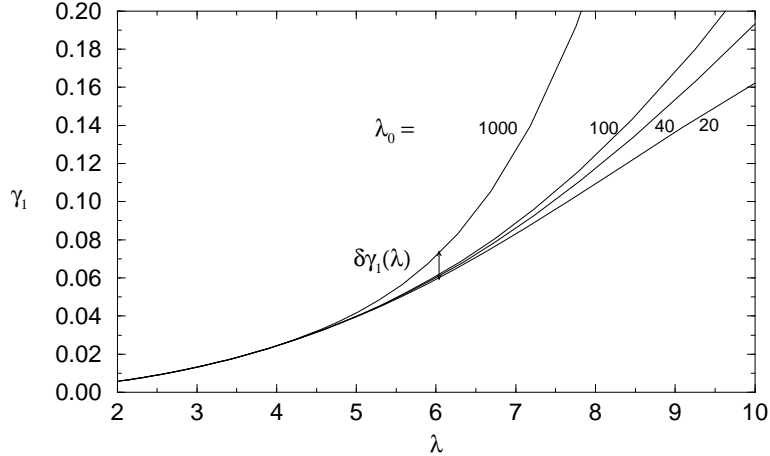}\hfill
\vskip -3cm
\caption{Trajectories for $\Lambda<\!\!<\Lambda_0$, starting from $\gamma_1(\Lambda_0)=0$,
and various values $\lambda(\Lambda_0)=\lambda_0$.}
\end{figure}
As can be seen from the curves, the irrelevent coupling $\gamma_1$
decays into what appears to be a well defined Renormalised Trajectory
for $\lambda\ltap4$.
(Actually this term is a misnomer here,  as we discuss in the
next subsection.)
In this regime, we may solve the equations (\ref{lams},\ref{gams})
perturbatively as follows. We recognize from (\ref{lams}) that $\beta$
is at least $O(\lambda^2)$, which implies from (\ref{gams}) that
$\gamma_1=\lambda^2/(24\pi^3)+O(\lambda^3)$. But given this information,
(\ref{lams}) yields the first two orders, $O(\lambda^2)$ and
$O(\lambda^3)$, in the $\beta$ function. Substituting these results in 
(\ref{gams}) yields the $O(\lambda^3)$ part of $\gamma_1$ and so on:\cite{erg}
\begin{eqnarray}
\label{gas}
\gamma_1 &&={1\over24\pi^3}\lambda^2+{1\over96\pi^5}\lambda^3+\cdots\\
\label{bas}
\beta &&={3\lambda^2\over(4\pi)^2}
+{8\over\pi}{\lambda^3\over(4\pi)^4}+\cdots
\end{eqnarray}
Similarly, the perturbative solution for the full theory 
may be worked out directly in renormalised terms.\cite{Bon}\cite{wie}

\subsection{Triviality and renormalons}
Since we can in this way compute directly the RT, without having to worry
about constructing a bare action and bare couplings,
does this mean that after all, an interacting continuum
limit for four dimensional $\lambda\phi^4$ theory exists? The answer is
of course negative. But it is instructive to understand why. 
 Firstly, one must understand that the solution
(\ref{gas},\ref{bas}) (and the equivalent
perturbative solution of the full theory) does {\sl not} parametrize the 
RT.\footnote{contrary to statements in the literature\cite{wie}}
The coupling $\lambda$ being marginally irrelevant,
 sinks back into the Gaussian fixed point $\lambda=\gamma_1=0$; there is
actually
only one direction out of the critical surface\footnote{here neglected}
 along ${\cal O}\sim \phi^2$ (plus quantum corrections),
corresponding to a RT which describes
a massive but non-interacting, i.e. trivial, theory.

Nevertheless, the trajectory  (\ref{gas},\ref{bas})  corresponds to an 
apparently unique trajectory within the critical surface
and one might wonder whether this is provides a clue to
a non-perturbative continuum limit for scalar
theory. Suppose for example, that there exists a non-perturbative
ultraviolet  fixed point.\cite{kogwil} Then there would
indeed be a unique trajectory, lieing in the critical surface,
 from this fixed point to
the Gaussian fixed point, as
sketched in fig.5. 
\begin{figure}[ht] 
\epsfxsize=0.6\hsize
\hfil\epsfbox{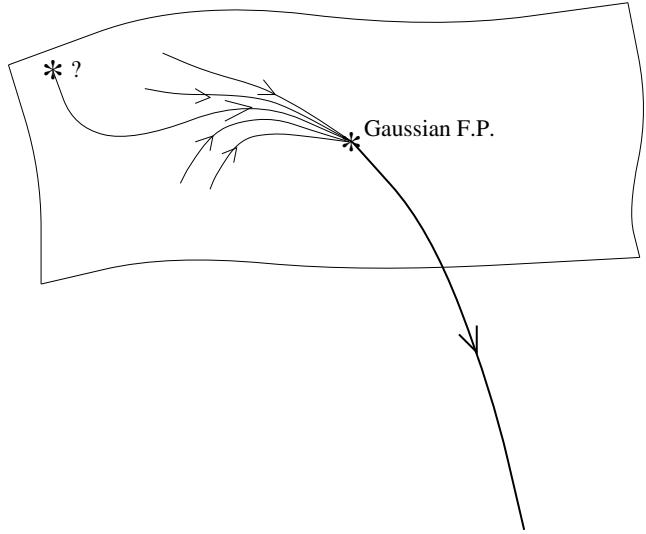}\hfill
\caption{A nontrivial fixed point would define
 a unique trajectory in the critical surface.}
\end{figure}
The new fixed point would then provide the
basis for a continuum limit, whose infrared behaviour could still be
 controlled by
the Gaussian fixed point, by tuning the bare action
along the relevant directions
out of the new fixed point. However this scenario is also false.
As we will see later there are no other fixed points (at
least within the local potential approximation). And importantly, 
the apparent uniqueness of (\ref{gas},\ref{bas}) is illusory.
Clearly, one can readily generate very high orders of perturbation 
theory in  (\ref{gas},\ref{bas}). As hinted
at the end of the introduction in ref.\cite{erg},
 the resulting series is divergent and 
non-Borel summable, which means that the series does not in itself 
determine a unique trajectory. 
 This is a consequence 
of ultraviolet renormalons.\cite{zinn}
The ambiguity can be determined by solving for a linearised perturbation
$\delta\gamma_1(\lambda)$ to (\ref{gas},\ref{bas}) in (\ref{lams},\ref{gams})
(indicated on fig.4). Neglecting multiplicative
perturbative corrections, we find
\begin{equation}
\label{uvren}
\delta\gamma_1(\lambda)\propto \lambda^{\tau}{\rm e}^{-1/
(\beta_1\lambda)}\quad,
\end{equation}
where the one-loop beta-function coefficient $\beta_1=3/(4\pi)^2$, and in
this approximation $\tau={2/3}-{8/(9\pi)}$.
(It is important to recognize that nothing about the Gaussian
fixed point determines the proportionality
constant here: all $\lambda$ derivatives of $\delta\gamma_1$ are zero
at the Gaussian fixed point, so the fact that the trajectory points
uniquely along the marginal direction here is insufficient to rule out
such perturbations.)

Integrating the $\beta$ function (\ref{bas}) shows that (\ref{uvren})
is nothing but the irrelevant perturbation $\delta\gamma_1\sim \Lambda/
\Lambda_0$ (plus quantum corrections) rewritten in terms of $\lambda$,
as again should be expected.\footnote{The peculiar scaling dimension
is again a consequence of sharp cutoffs.\cite{erg}\cite{truncm}
For smooth cutoffs only even dimensions would appear.}
 On the other hand, if (\ref{uvren})
delineates the leading singularity of the non-perturbative part
of $\gamma_1(\lambda)$ in the
complex $\lambda$-plane, then for high orders in $\lambda$,
by Cauchy's theorem:
\begin{equation}
\gamma_1(\lambda)\sim A\lambda\int_0^\infty\!\!\!dg \,
{g^{\tau-1}\over g-\lambda}\,{\rm e}^{-1/(\beta_1 g)}\quad,
\end{equation}
for some coefficient $A$.
Expanding this as a perturbation series in $\lambda$, 
we obtain
\begin{equation}
\label{lo}
\gamma_1(\lambda)\sim A\sum_m (\beta_1\lambda)^m\Gamma(m-\tau)\quad.
\end{equation}
This indeed matches
the large order behaviour of the perturbation series (\ref{gas}).
This is the phenomenon of resurgence in asymptotic series.
Using the large order terms of (\ref{gas}), we can
determine a value for $A$ which corresponds to 
the simplest Borel contour. However, this yields a complex
renormalon coefficient for (\ref{uvren}).

We see that the way renormalons appear, at least the ultraviolet ones, is
particularly straightforward and intuitive in this framework, as might well
have been expected. 
The same behaviour will appear in the full theory, with an infinite
number of ultraviolet
renormalon contributions, 
one from each irrelevant operator. 

\section{Approximations}
After this diversion into the world of perturbation theory, we return to
our purely non-perturbative discussion. We have seen that all continuum
limits follow directly from the fixed points $S_*[\phi]$ and marginally
relevant,
and relevant eigenoperators ${\cal O}_i[\phi]$, and their associated
RG eigenvalues $\lambda_i\ge0$. 

Since any approximation which preserves the fact that the RG
flow equations are non-linear will continue to have this structure
of fixed points and perturbations, {\sl any such approximation will 
preserve the existence of continuum limits, and thus renormalisability and
self similar flows}: $S[\phi]({\underline g},\gamma)$. For a 
particular fixed point, it is only necessary to ensure that its 
desired qualitative features (e.g. the number of relevant and marginally
relevant eigenperturbations) are reproduced in the approximation.

\subsection{Truncations}
The simplest form of approximation is to truncate the effective action
$S_\Lambda$ so that it contains just a few operators. The $\Lambda$
dependent coefficients of these operators then
have flow equations determined by equating coefficients on the left
and right hand side of (\ref{Pfl}) [or (\ref{Gfl})], 
after rejecting from the right
hand side of (\ref{Pfl}) all terms that do not `fit' into this set
(this being the approximation). The difficulty with this approximation
is that it inevitably
results 
in a truncated
expansion in powers of the field $\varphi$ (about some point),
 which can only be sensible if
the field $\varphi$ does not fluctuate very much, which is the
same as saying that it is close to mean
field,\cite{rg96} i.e. in a setting in which weak coupling perturbation
theory\footnote{or in some settings, 
large $N$ approximations\cite{AokN,wegho,lN,LN}}
is anyway valid.
This is precisely the opposite regime
from the truly non-perturbative one that concerns us here. 
(Generically in this regime, one finds that higher orders of truncation 
cease to converge and reliability is lost since many spurious fixed
points are generated.\cite{trunc})

It is hard to see how this, admittedly qualitative, argument can fail
in practice. However, truncations of powers of the field $\phi$
around the minimum of the effective potential in scalar field theory
appear to provide an 
exception.\cite{AokN,tetwet,goodtr,zak} 
At high orders of truncation it is
possible to obtain as much as  9 digits accuracy,\cite{AokN2}
 before succumbing to
the generic  pattern 
for finite truncations  as 
outlined above.\cite{A,trunc}
Recalling the above qualitative argument, 
the success of these truncations suggests to me
that some sort of perturbation theory may in fact be 
applicable to this case in practice.

\subsection{Derivative expansion}
A less severe, more natural, and more accurate 
expansion, closely allied to the 
successful truncations in real space RG of 
spin systems,\cite{rg96,zak} is rather to perform a `short distance
expansion'\cite{rg96,truncm}
 of the effective action $S_\Lambda$, which for smooth cutoff
profiles corresponds to a 
derivative expansion:\cite{deriv,eqs,gol,twod,ball,on,Jor}
\begin{equation}
\label{dexp}
S_\Lambda\sim\int\!\!d^D\!x\,\left\{V(\phi,\Lambda)+
\half(\partial_\mu\phi)^2 K(\phi,\Lambda)
+O(\partial^4)\right\}\quad.
\end{equation}
[In the $N\ne1$ component case there is a second $O(\partial^2)$ term:
$\half(\phi^a\partial_\mu\phi^a)^2 Z(\phi,\Lambda)$.]

The simplest such approximation is the so-called 
Local Potential Approximation (LPA),
introduced by Nicoll, Chang and Stanley:\cite{nico}
\begin{equation}
\label{LPA}
S_\Lambda\sim\int\!\!d^D\!x\,\left\{V(\varphi,\Lambda)+{1\over2}
(\partial_\mu\varphi)^2
\right\}\quad.
\end{equation}
It has since
been rediscovered by  
many authors,\cite{truncm}
notably Hasenfratz and Hasenfratz.\cite{hashas}
As a concrete example, consider the case of sharp cutoff. The flow equations
may be shown to reduce to\cite{nico,hashas,erg,trunc,truncm}
\begin{equation}
\label{Vfl}
{\partial\over\partial t}V(\varphi,t)+d\,\varphi V'
-DV=\ln(1+V'')\quad,
\end{equation}
where $\prime\equiv{\partial\over\partial\varphi}$,
$t=\ln(\mu/\Lambda)$ and $d=\half(D-2)$. In the $N$ component
case, the right hand side of (\ref{Vfl})
has an extra term $+(N-1)\ln(1+V'/\phi)$
(and $\phi$ stands for the length of the $\phi^a$ vector),
however for the most part it will be sufficient to consider
a single component scalar.
Actually, the  $N=\infty$ case 
was already derived by Wegner and
Houghton in their paper introducing the sharp-cutoff flow 
equation.\cite{wegho}
In this limit the LPA is effectively
exact.\cite{lN} It follows from our earlier discussion\cite{lN,eqs,erg} 
that $V$ is an approximation to the
Legendre effective potential in the limit $\Lambda\to0$, i.e. $t\to\infty$.

\section{Fixed points}
We have not yet addressed the question as to how the fixed points and eigenperturbations
 are determined within the exact RG.
We know from other methods,
including experiment, that there are generically a discrete set of RG
fixed point solutions and a discrete set of eigenperturbations, but here
these quantities are determined from functional {\sl differential} equations
(\ref{Sfl}) or (\ref{Gfl})
with thus, apparently, a continuum of solutions.
The derivative expansion approximations,
also have the property that the RG flow equations are differential equations
with a continuum of solutions. 
We will show how one finds nevertheless (generically) only
a discrete set of acceptable solutions within these approximations.
Since this is true to all orders of the derivative expansion,
we assume that there are only a discrete set of acceptable solutions
of the exact RG, for the same reasons.

To begin with, we need only consider the LPA, and for concreteness we will
take the example of (\ref{Vfl}). In this case the fixed point potential $V_*$
satisfies
\begin{equation}
\label{V*}
d\,\varphi V'_*(\varphi)
-DV_*(\varphi)=\ln(1+V''_*)\quad.
\end{equation}
This equation has indeed a continuum of solutions, in fact a 
continuous two-parameter set. However, generically
all but a countable number
of these solutions are singular!\cite{trunc,hh}
 ($D=2$ dimensions is an exception.\cite{twod})
To illustrate this, let us choose $V_*'(0)=0$. (This is anyway necessary
if $N\ne1$.\cite{zak}) This fixes one boundary condition at $\phi=0$.
If we choose some numerical value for $V_*(0)$,
we then have the required two boundary conditions and can
 numerically integrate
(\ref{V*}) out to positive $\phi$.
 We find that almost without exception
a singularity is encountered at some critical value of the field $\phi=\phi_c$.
(At a very basic level this is indicated by the failure of the numerical
routine near this point, although it is possible to solve analytically
for the singularity and use this information to do much
better.\cite{trunc,deriv,twod}) The value of $\phi_c$ depends on $V_*(0)$.
In fig.6 we plot the results for two examples. 
\begin{figure}[ht] 
\vskip -1.5cm 
\epsfxsize=0.7\hsize
\hfil\epsfbox{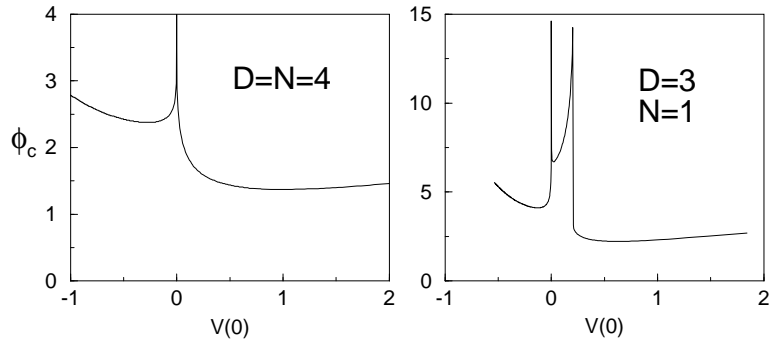}\hskip 1cm\hfill
\vskip -2cm
\caption{Plots of $\phi_c$ against $V(0)$ for $D=N=4$, and $D=3$ and
$N=1$.}
\end{figure}
The first graph  is a plot for the case $D=4$ and $N=4$ (the
Higgs field in the Standard Model). We see that only
the (trivial) Gaussian fixed point solution $V(\phi)\equiv0$ exists for 
all values
of the field. If the same is done for the case $D=3$ and $N=1$, we get the
second graph. In this case there is also one non-trivial non-singular
solution, corresponding to the famous Wilson-Fisher fixed point
(Ising model universality class). 

Note that this straightforward numerical procedure corresponds 
nevertheless, within the LPA, to an exhaustive search 
for continuum limits in the
entire infinite dimensional space of all possible potentials $V(\phi)$.
Similar entire searches are possible at higher orders of the derivative
expansion.\cite{twod,rg96} 
Clearly this is much more than is possible with other methods!

For large field $\varphi$ the only consistent behaviour (with $D>2$) for
the fixed point potential in (\ref{V*}) is 
\begin{equation}
\label{V*as}
V_*(\varphi)\sim A \varphi^{D/d}\quad,
\end{equation}
where $A$ is a constant determined by
the equations. 
This simply solves the left hand side of (\ref{V*}), 
these terms arising from purely dimensional considerations,
and neglects
the right hand side of the flow equation -- which encodes the
quantum corrections. Or in other words,
(\ref{V*as}) is precisely what would be expected by dimensions
(since $V$'s mass-dimension is $D$ and $\varphi$'s is $d$)
providing only that any dependence on $\Lambda$, and thus
the remaining quantum corrections, can
be neglected. Requiring the form (\ref{V*as}) to hold for both
$\varphi\to\infty$ and $\varphi\to-\infty$, provides 
the necessary two boundary
conditions for the second order ordinary differential equation (\ref{V*}),
so we should indeed generally
expect at most a discrete set of globally non-singular solutions.
These considerations generalise to any order of the derivative expansion,
and indeed we thus expect them to hold also for the exact RG.
(There is one
modification: beyond LPA, $d={1\over2}(D-2+\eta)$, where $\eta$ is
the anomalous dimension at the fixed point. We will discuss this later.)

\section{Eigenoperators}
Now consider the determination of the eigenoperators. For this
we perturb away from the fixed point:
\begin{equation}
\label{Vv}
V(\phi,t)=V_*(\phi)+v(\phi,t)\quad.
\end{equation}
For $v<\!\!< V_*$, we can expand  (\ref{Vfl}) to first order in $v$. Then
by separation of variables,
\begin{equation}
\label{vfl}
v(\phi,t)=\alpha \, {\rm e}^{\lambda t} u(\varphi)\quad,
\end{equation}
as in (\ref{RTbc}), where $\alpha$
is a small parameter and $u(\phi)$ is some (normalised) solution of
\begin{equation}
\label{u}
\lambda u + d\,\varphi u'-Du={u''\over1+V''_*}\quad.
\end{equation}
We again have a two parameter continuum of solutions, but in this case
each one is guaranteed globally
well defined since (\ref{u}) is linear,
and this is true
for {\sl every} value of $\lambda$. How can this be squared with the
expectation of only a 
discrete spectrum of such operators? Firstly, one of the parameters corresponds
just to the overall normalisation. The second parameter and $\lambda$ are
fixed however, for a much more subtle reason:
{\sl only the
discrete set of normalised solutions for $u(\varphi)$ that behave as a power of
$\varphi$ for large field, can be associated with a corresponding 
renormalised coupling $g(t)$ and thus the 
universal self-similar flow (\ref{ssfl})
which is characteristic of the continuum limit}.\cite{hh,eqs,zak}

Indeed we see from (\ref{u}) and (\ref{V*as}), that those solutions
that behave as a power for large $\varphi$ must do so as
\begin{equation}
\label{upo}
u(\varphi)\sim  \varphi^{(D-\lambda)/d}\quad,
\end{equation}
this being again the required power to balance scaling dimensions 
(with $[g(t)]=\lambda$) if the remaining quantum corrections may
be neglected in this regime. Once again for $\varphi\to\pm\infty$, this
supplies two boundary conditions, but 
this time, since (\ref{u}) is linear, this
overdetermines the equations, and generically
allows only certain quantized values of $\lambda$.

On the other hand if $u$ does not behave as a power of $\varphi$ for
large $\varphi$, then from (\ref{u}) and (\ref{V*as}), we obtain
that instead for large $\phi$
\begin{equation}
\label{unonpo}
u(\varphi)\sim\exp\left\{A(D-d)\varphi^{D/d}\right\}\quad.
\end{equation}
(The precise form of the large $\varphi$ dependence of the 
non-power-law perturbations
depends on non-universal details including the level
of derivative expansion approximation used, if any.
Universally however, the non-power-law perturbations grow faster than a power,
and this is all we will really need.)

To investigate whether these perturbations are associated with renormalised
couplings, we must follow the evolution of the perturbed action (\ref{Vv})
for a small but {\sl finite} $v$.
Starting, say, with (\ref{vfl}) at $t=0$ as a boundary
condition: 
\begin{equation}
\label{vbc}
v(\phi,0)=\alpha u(\phi)\quad,
\end{equation}
we must show that the evolved
solution $v(\phi,\alpha,t)$ can be expressed as a self-similar flow,
$v(\phi,g(t))$, for some renormalised coupling $g$.

Now by (\ref{V*as}), for all perturbations behaving
as (\ref{unonpo}), or (\ref{upo}) when $\lambda\le0$, 
there is potentially a problem because for any
small but finite $\alpha$, there will always be a $\phi$ large enough
where we cannot treat $v(\phi,t)$ as small compared to $V_*$.
Therefore the linearised solution
(\ref{vfl}), and indeed more generally (\ref{RTbc}),
 needs reexamining in this regime. Fortunately, 
in the large $\varphi$ regime, we may solve (\ref{Vfl}) non-perturbatively
(and thus without making any assumption on the size of $v/V_*$). 
This is because, just as before,
we may neglect the quantum corrections in this regime.
These are given by the right hand side of (\ref{Vfl}),
and thus $V(\varphi,t)$ follows 
mean-field-like evolution:
\begin{equation}
\label{mf}
V(\varphi,t)\sim\, {\rm e}^{Dt} V(\varphi\,{\rm e}^{-dt},0)\quad.
\end{equation}
Applying this to (\ref{Vv}) and (\ref{vbc}), we see that in the
large $\phi$ regime these perturbations evolve as
\begin{equation}
\label{vnonpert}
v(\phi,\alpha,t)\sim\alpha\,{\rm e}^{Dt}u({\rm e}^{-dt}\phi)\quad.
\end{equation}
For the power-law perturbations, (\ref{upo}) then implies 
\begin{equation}
\label{vren}
v(\phi,\alpha,t)\sim\alpha\,{\rm e}^{\lambda t}\varphi^{(D-\lambda)/d}\quad,
\end{equation}
which is indeed self-similar with renormalised coupling $g(t)\sim
\alpha\,{\rm e}^{\lambda t}$.  (We also see that the linearised solution
(\ref{vfl}) is afterall still valid in this regime.)
On the other hand, using  (\ref{unonpo}) 
we see that in the large $\phi$ regime the non-power-law 
perturbations behave as
\begin{equation}
\label{vnonren}
v(\phi,\alpha,t)\sim\alpha\,
\exp\left\{Dt+A(D-d)\,{\rm e}^{-Dt}\varphi^{D/d}\right\}\quad.
\end{equation}
This cannot be rewritten as a universal self-similar flow $v(\phi,g(t))$.

Use of the `relevant' non-power-law  perturbations
as the basis for a Renormalised 
Trajectory (\ref{RTbc}) would not
be valid: on the one hand the 
 linearised solution (\ref{vfl}) is not valid when $\phi$ is large,
and on the other hand
the full solution (\ref{vnonren}) 
does not fall back into the fixed point
as $t\to-\infty$ ($\Lambda\to\infty)$.
Indeed this limit does not even exist.
By Sturm-Liouville analysis,\cite{eqs} one can further show 
from (\ref{vnonren}) that 
these non-power-law perturbations collapse, on increasing $t$, into an infinite
sum of the quantized power-law perturbations, and thus the 
non-power-law eigenperturbations are entirely irrelevant for 
continuum physics.

\section{Anomalous dimension and reparametrization invariance}
So far we have been ignoring the determination of $\eta$,
the fields anomalous
dimension.  Recall that this arises from the anomalous scaling of
the field necessary in general to ensure that the kinetic term
$\sim(\partial_\mu\phi)^2$ of the effective action is conventionally
normalised, and more importantly to ensure that effective
actions are finite and do
actually achieve fixed points at scale invariant continuum
limits.  In the LPA, the lowest order of
the derivative expansion, all momentum dependent corrections to the
effective action are thrown away and thus $\eta$ is always zero
in this approximation.  Beyond LPA, and in the exact RG, we must scale
out the field according to its full scaling dimension (as described at
the end of sect.2): $\phi\sim \Lambda^d$, where now
\begin{equation}
d=\half(D-2+\eta)\quad.
\end{equation}
(In this section, it will be sufficient for us to concentrate
on the behaviour at fixed points and thus take $\eta$ as independent
of $\Lambda$.)
The flow equations at higher order in the derivative  expansion 
(\ref{dexp}), now take the generic form
\begin{eqnarray}
\label{Vfl2}
{\partial\over\partial t}V(\varphi,t)+d\,\varphi V'
-DV &&= \cdots \\
\nonumber
{\partial\over\partial t}K(\varphi,t)+\eta\, K &&=\cdots
\end{eqnarray}
(and so on for other coefficient functions).
Here, the terms on the left of the equation arise once again purely from
the assignment of scaling dimensions. The terms on the right of the
equations arise from the right hand side of the exact flow equations
(\ref{Sfl},\ref{Gfl}), and as a consequence of the structure of
(\ref{Sfl},\ref{Gfl}), are non-linear and reduce to
 second order differential equations.  As an example, the $O(\partial^2)$
approximation of the Legendre flow equations (\ref{Gfl}) in $D=3$
dimensions (with a particular form of cutoff $C_{IR}$ which we will
discuss later) yields\cite{deriv}
\begin{eqnarray}
&&\label{Vflod2}\phantom{\hbox{and}\hskip 1cm}
{\partial V\over\partial t}+{1\over2}(1+\eta)\phi V'-3V=
-{1-\eta/4\over\sqrt{K}\sqrt{V''+2\sqrt{K}} } \\
&&\label{Kflo}\hbox{and}\hskip 1cm
{\partial K\over \partial t}+{1\over2}(1+\eta)\phi K' +\eta K=
\left(1-{\eta\over4}\right)\Biggl\{ 
{1\over48}{24KK''-19(K')^2\over K^{3/2}(V''+2\sqrt{K})^{3/2}} \\
&&\nonumber
-{1\over48}{58V'''K'\sqrt{K}+57(K')^2+(V''')^2K\over K(V''+2\sqrt{K})^{5/2}}
+{5\over12}{(V''')^2K+2V'''K'\sqrt{K}+(K')^2\over\sqrt{K}(V''+2\sqrt{K})^{7/2}}
\Biggr\}\ \ .
\end{eqnarray}
(The perceptive reader will note that the right hand sides of (\ref{Kflo})
actually contain $V'''$'s. However, differentiating (\ref{Vflod2}) yields
an equation for $V'''$ in terms of expressions with lower derivatives,
which when substituted into (\ref{Kflo}) reduces this to second order
as claimed.)

Once again, one finds that the fixed point equations generically
have at most a discrete set of non-singular solutions,
and that this can be understood by studying the large $\phi$
behaviour.\cite{deriv,twod,rg96,zak} Indeed we find for large $\phi$ that
the right hand sides of  (\ref{Vfl2}) are subleading for the
power-law behaviour that solves the left hand sides: 
\begin{equation}
\label{infbcs}
V\sim A_V\,\phi^{D/d}\quad\quad\quad\quad K\sim A_K\,\phi^{-\eta/d}\quad,
\end{equation} 
etc, 
where $A_V$ and $A_K$ are constants that get determined by the equations.
Once again this large $\phi$
behaviour is precisely to be expected by scaling dimensions, providing
only that any dependence on $\Lambda$, and thus the remaining quantum
corrections, can be neglected in this regime.
Requiring that these hold for both $\phi\to\infty$ and
$\phi\to-\infty$ provides the required two boundary conditions for each
coefficient function, so that we should expect generically at most a
discrete set of such globally non-singular solutions.  However, this time
this is true for {\sl each} $\eta$. How then, does $\eta$ get determined?

In the original partition function (\ref{zorig}), physics does not depend
on the normalization of $\phi$, i.e. we may substitute 
\begin{equation}\label{parami}\phi\mapsto\Omega\phi\end{equation} 
in the action {\sl while leaving the $J\cdot\phi$ term alone}, or
equivalently mapping only $J\mapsto J/\Omega$ and leaving the field alone. 
 More generally
we may exchange $\phi$ for $\phi$ plus any local expression in $\phi$.
This is the ``equivalence theorem''.\cite{itz}
The continuous RG has a similar 
``reparametrization'' symmetry\cite{wegred,red,rie}
but  in general it is given by a complicated functional 
integral transform.\cite{rie} 
Thus {\sl all fixed points} appear as 
lines of {\sl equivalent} fixed points
generated by
an, in general 
complicated, exactly marginal perturbation.
This means that the boundary conditions
(\ref{infbcs}), which as we have seen are 
already sufficient to constrain the
fixed point solutions down to a discrete set, actually overconstrain the
solution space and lead to quantization of $\eta$
(in a similar way to the linear equations for eigenperturbations).
Equivalently, note that
 we have the freedom to choose an
extra boundary condition, a normalization condition,  e.g. by
requiring a
conventionally normalized kinetic term $K_*(0)=1$. For such representatives
(of the equivalence classes under reparametrization)
the fixed point equations are overconstrained, leading to quantization
of $\eta$.
In this way,  the
reparametrization invariance turns the fixed point equations
into {\sl non-linear eigenvalue equations} for $\eta$.\cite{wegred}

There is a problem however for approximations:
 the derivative expansion, indeed any truncation of the momentum or
field dependence, generally breaks
the reparametrization invariance, with the result that $\eta$
(and other universal quantities such as the RG eigenvalues) 
depend on some unphysical parameter such as $K_*(0)$.\cite{gol}
Note that even without the reparametrization invariance we could choose
to fix e.g. $K_*(0)=1$. The problem is that, if the invariance is
broken,  different values for $\eta$ and other universal quantities  will
be obtained for different normalisations $K_*(0)\ne1$, violating the equivalence
theorem.\footnote{Some techniques
have been developed to suggest a `best' choice amongst
 the one-parameter solution
set for a given fixed point in these cases.\cite{gol,Jor}
Similar practical issues exist, in general, 
for the choice of cutoff function $C_{IR}$: universal quantities are 
independent of the detailed choice in the exact RG flow equations, but
in general for approximations the results do depend on the choice
and this raises the issue of choosing a `best' function out of some 
set. \cite{deriv,ball,Jor}}
Depending on how the equations are parametrized,
this ambiguity can be shifted around into different quantities but it is
always there if reparametrization invariance is absent.
In this sense, the problem can be quite subtle to spot,
and has been missed in a number of recent works.
But the general question any practitioner must ask is if {\sl all} fixed point
solutions are investigated, where care is taken to ensure that
{\sl only} arbitrary normalisations are set in the solutions,
are the results independent of  these normalisations? Evidently this
is only true if there exists some underlying reparametrization invariance.

We will demonstrate the existence of reparametrization invariance in a simple
example. Recall from sect.2, that the Wilsonian effective action is given by
\begin{equation}
S^{tot}_\Lambda[\phi]=\half\phi\cdot\Delta^{-1}_{UV}\cdot\phi
+S_\Lambda[\phi]\quad,
\end{equation}
where $\Delta_{UV}=C_{UV}(q,\Lambda)/q^2$,
and the effective interaction part $S_\Lambda$ is governed by the flow equation 
(\ref{Sfl}). Clearly the Gaussian fixed point is given by $S_\Lambda=0$. But this
is not the only Gaussian fixed point solution! Since (\ref{Sfl}) is given in terms of 
unscaled variables (the change to dimensionless
 variables at the end of sect.2 having not yet been
done), it is easier for this example to work with the unscaled variables. 
We specialize to $C_{UV}\equiv C_{UV}(q^2/\Lambda^2)$.\footnote{
This is required by the canonical scaling of the Gaussian fixed point
and Lorentz invariance.}
If we take as ansatz
\begin{equation}
\label{ansatz}
S_\Lambda=\half\phi\cdot q^2 z(q^2/\Lambda^2)\cdot\phi\quad,
\end{equation}
then any such solution will be a fixed point after the change to dimensionless
variables $\phi\mapsto\Lambda^{(D-2)/2}\phi$, $q\mapsto\Lambda q$.
Substituting (\ref{ansatz}) into (\ref{Sfl}), we find that $z'= 
z^2C'_{UV}$ (prime being differentiation
with respect to its argument). This has general solution $z=1/(a-C_{UV})$, 
for some integration
constant $a>1$.\footnote{The restriction on $a$ arises 
because $z$ must be nonsingular.
We assume that $C_{UV}\le1$. Such is 
the case if $C_{UV}$ is monotonic for example.} 
Thus the general fixed point solution takes the form
\begin{equation}
\label{Sans}
S^{tot}_\Lambda={1\over2}\phi\cdot{a\,
q^2\over C_{UV}(a-C_{UV})}\cdot\phi\quad.
\end{equation}
This line of fixed points, parametrized by $a$, 
is a line of {\sl equivalent} fixed points
as can be confirmed by checking that the 
spectrum of RG eigenvalues (and thus in particular the critical exponents) 
is still that of the
Gaussian fixed point.\cite{Jor} 
The reparametrization invariance that maps between 
one representative and another clearly has a 
complicated momentum dependence
for general $C_{UV}$.
Also, a derivative expansion of (\ref{Sans}) corresponds
to a Taylor expansion in $q^2$,
and thus clearly any derivative expansion (to finite order)
will destroy the equivalence of these
fixed points, and the reparametrization invariance, for general $C_{UV}$. 

If we take the cutoff sharp 
$C_{UV}=\theta(\Lambda-q)=\theta(1-q^2/\Lambda^2)$, then 
\begin{equation}
{C_{UV}(a-C_{UV})\over a}= \left({a-1\over a}\right) C_{UV}\quad.
\end{equation}
(as can easily be confirmed by comparing both sides for $q>\Lambda$ and
$q<\lambda$).
Since this is the inverse of the cutoff terms appearing in (\ref{Sans}), we see that 
in the sharp cutoff case the reparametrization invariance is simply the linear
momentum independent transformation (\ref{parami}). This can be shown to be
true in general for the flow equations themselves.\cite{truncm,rg96,Jor}
It is not possible however to use a 
derivative expansion in the case of a sharp cutoff;
instead
an expansion in `momentum scale' may be used.\cite{truncm}

In general, one may show that the flow equations (\ref{Sfl},\ref{Gfl}) enjoy a 
momentum independent and linearly realised reparametrization symmetry if and
only if either $C_{IR}=\theta(q-\Lambda)$ i.e. the sharp case just
discussed, or $C_{IR}^{-1}=1+(\Lambda^2/q^2)^k$ ---
a power-law cutoff.\footnote{$k$ should be chosen to be
an integer. A function
of the scale can be included in front of the 
monomial.\cite{deriv,twod}}\cite{deriv,twod,rg96}

The power-law case can be used with derivative expansions, but it only
regulates the Legendre flow equation (\ref{Gfl}). 
This is the cutoff that was used 
(with $k=2$) to produce the examples (\ref{Vflod2},\ref{Kflo}).
The reparametrization invariance takes the form
$\phi\mapsto\phi\Omega^{k+(D-2)/2}$, $q\mapsto q\Omega$.
In the examples (\ref{Vflod2},\ref{Kflo}) this implies that the
equations are invariant under 
$\phi\mapsto\Omega^{5/2}\phi$, $V\mapsto\Omega^3V$, $K\mapsto\Omega^{-4}K$.
It is easy to check that (\ref{Vflod2},\ref{Kflo}) are indeed invariant
under this symmetry.

\section*{Acknowledgements}
It is a pleasure to thank Tsuneo Suzuki and the rest of the organizing
committee for the invitation to speak at this pleasurable and well 
run international seminar, several participants -- particularly
Ken-Ichi Aoki, Haruhiko Terao and their students, and Peter Hasenfratz,
for enjoyable and informative conversations, and finally the SERC/PPARC for
financial support through an Advanced Fellowship.



\vfill\eject

\end{document}